\documentstyle[12pt]{article}

\catcode`\@=11
\long\def\@makefntext#1{
\protect\noindent \hbox to 3.2pt {\hskip-.9pt
$^{{\ninerm\@thefnmark}}$\hfil}#1\hfill}		%CAN BE USED

 \def\@makefnmark{\hbox to 0pt{$^{\@thefnmark}$\hss}}  %ORIGINAL

\def\ps@myheadings{\let\@mkboth\@gobbletwo
\def\@oddhead{\hbox{}
\rightmark\hfil\ninerm\thepage}
\def\@oddfoot{}\def\@evenhead{\ninerm\thepage\hfil
\leftmark\hbox{}}\def\@evenfoot{}
\def\sectionmark##1{}\def\subsectionmark##1{}}

\newcounter{sectionc}\newcounter{subsectionc}\newcounter{subsubsectionc}
\renewcommand{\section}[1] {\vspace{0.6cm}\addtocounter{sectionc}{1}
\setcounter{subsectionc}{0}\setcounter{subsubsectionc}{0}\noindent
	{\bf\thesectionc. #1}\par\vspace{0.4cm}}
\renewcommand{\subsection}[1] {\vspace{0.6cm}\addtocounter{subsectionc}{1}
	\setcounter{subsubsectionc}{0}\noindent
	{\it\thesectionc.\thesubsectionc. #1}\par\vspace{0.4cm}}
\renewcommand{\subsubsection}[1]
{\vspace{0.6cm}\addtocounter{subsubsectionc}{1}
	\noindent {\rm\thesectionc.\thesubsectionc.\thesubsubsectionc.
	#1}\par\vspace{0.4cm}}

\newcounter{appendixc}
\newcounter{subappendixc}[appendixc]
\newcounter{subsubappendixc}[subappendixc]

\renewcommand{\appendix}[1] {\vspace{0.6cm}
        \refstepcounter{appendixc}
        \setcounter{figure}{0}
        \setcounter{table}{0}
        \setcounter{equation}{0}
        \renewcommand{\thefigure}{\Alph{appendixc}.\arabic{figure}}
        \renewcommand{\thetable}{\Alph{appendixc}.\arabic{table}}
        \renewcommand{\theappendixc}{\Alph{appendixc}}
        \renewcommand{\theequation}{\Alph{appendixc}.\arabic{equation}}
        \noindent{\bf Appendix \theappendixc #1}\par\vspace{0.4cm}}

\def\abstracts#1{{
	\centering{\begin{minipage}{30pc}\tenrm\baselineskip=12pt\noindent
	\centerline{\tenrm ABSTRACT}\vspace{0.3cm}
	\parindent=0pt #1
	\end{minipage}}\par}}

\renewenvironment{thebibliography}[1]
	{\begin{list}{\arabic{enumi}.}
	{\usecounter{enumi}\setlength{\parsep}{0pt}
\setlength{\leftmargin 1.25cm}{\rightmargin 0pt}
	 \setlength{\itemsep}{0pt} \settowidth
	{\labelwidth}{#1.}\sloppy}}{\end{list}}

\newcounter{itemlistc}
\newcounter{romanlistc}
\newcounter{alphlistc}
\newcounter{arabiclistc}
\newenvironment{itemlist}
    	{\setcounter{itemlistc}{0}
	 \begin{list}{$\bullet$}
	{\usecounter{itemlistc}
	 \setlength{\parsep}{0pt}
	 \setlength{\itemsep}{0pt}}}{\end{list}}

\newenvironment{romanlist}
	{\setcounter{romanlistc}{0}
	 \begin{list}{$($\roman{romanlistc}$)$}
	{\usecounter{romanlistc}
	 \setlength{\parsep}{0pt}
	 \setlength{\itemsep}{0pt}}}{\end{list}}

\newcommand{\fcaption}[1]{
        \refstepcounter{figure}
        \setbox\@tempboxa = \hbox{\tenrm Fig.~\thefigure. #1}
        \ifdim \wd\@tempboxa > 6in
           {\begin{center}
        \parbox{6in}{\tenrm\baselineskip=12pt Fig.~\thefigure. #1}
            \end{center}}
        \else
             {\begin{center}
             {\tenrm Fig.~\thefigure. #1}
              \end{center}}
        \fi}

\newcommand{\tcaption}[1]{
        \refstepcounter{table}
        \setbox\@tempboxa = \hbox{\tenrm Table~\thetable. #1}
        \ifdim \wd\@tempboxa > 6in
           {\begin{center}
        \parbox{6in}{\tenrm\baselineskip=12pt Table~\thetable. #1}
            \end{center}}
        \else
             {\begin{center}
             {\tenrm Table~\thetable. #1}
              \end{center}}
        \fi}

\def\@citex[#1]#2{\if@filesw\immediate\write\@auxout
	{\string\citation{#2}}\fi
\def\@citea{}\@cite{\@for\@citeb:=#2\do
	{\@citea\def\@citea{,}\@ifundefined
	{b@\@citeb}{{\bf ?}\@warning
	{Citation `\@citeb' on page \thepage \space undefined}}
	{\csname b@\@citeb\endcsname}}}{#1}}

\newif\if@cghi
\def\cite{\@cghitrue\@ifnextchar [{\@tempswatrue
	\@citex}{\@tempswafalse\@citex[]}}
\def\citelow{\@cghifalse\@ifnextchar [{\@tempswatrue
	\@citex}{\@tempswafalse\@citex[]}}
\def\@cite#1#2{{$\null^{#1}$\if@tempswa\typeout
	{IJCGA warning: optional citation argument
	ignored: `#2'} \fi}}

\def\fnt#1#2{\footnotetext{\kern-.3em
	{$^{\mbox{\sevenrm #1}}$}{#2}}}

 1
 1
 1

\font\tenbf=cmbx10
\font\tenrm=cmr10
\font\tenit=cmti10

\font\ninerm=cmr9

\newcommand{\ds}{\mbox{$D^{\ast}$}}
\newcommand{\dss}{\mbox{$D^{\ast \ast}$}}

\begin{document}

\begin{flushright}
\footnotesize hep-ph/9410323
\end{flushright}
\vspace{0.4cm}
\centerline{\tenbf GLUONIC EXCITATIONS IN HEAVY MESONS AND}
\baselineskip=16pt
\centerline{\tenbf THEIR DECAYS BY FLUX-TUBE BREAKING}
\vspace{0.8cm}
\centerline{\tenrm PHILIP PAGE\footnote{Talk presented at {\it `Quark
Confinement and the Hadron Spectrum'}, 20-24 June 1994, Como, Italy
and based on work done in collaboration with Frank Close\cite{page}.}}
\baselineskip=13pt
\centerline{\tenit Theoretical Physics, University of Oxford,
1 Keble Road, Oxford OX1 3NP, UK}
\vspace{0.9cm}
\abstracts{We have examined decays of low-lying gluonic
excitations of mesons (hybrids) by chromo\-electric flux-tube breaking. An
analytical calculation of non-relativistic flux-tube model decay amplitudes
is performed in an harmonic oscillator approximation. Specific decay
signatures of all $J^{PC}$ charmonium hybrids are identified and the widths
predicted. We introduce a new selection rule which can be used to
understand the systematics of numerical decay calculations.}

\vfil
%\vspace{0.8cm}
\rm\baselineskip=14pt
\section{Mesons with an excited gluonic field}
\vspace{-0.2cm}
We define {\it charmonium hybrids} as charm-anticharm bound systems
(mesons) with an excitation of the gluonic degree of freedom. Unless
otherwise stated, we just refer to these systems as {\it `hybrids'}
from now on. They represent new
confined states of QCD beyond the quark model, and hence an {\it
important} experimental test of QCD. Interesting features of
hybrids are :

\begin{itemlist}

\item Their uniqueness as a bound systems with both `valent' fermions and
bosons.

\item Hybrids often have {\it exotic quantum numbers} not found in the
quark model, e.g. $J^{PC} = 0^{+-},1^{-+}$ and $2^{+-}$ for the lowest
lying hybrids in the flux-tube model\cite{isgur}, which facilitate easier
experimental detection of non-quark model states.

\item Hybrids are expected to have {\it masses} of $4.1
\pm 0.4$ GeV (literature average)\footnote{A mass of $\approx 4.3$ GeV has
been estimated with a numerical simulation in the flux-tube model\cite
{private}. The fact that these masses are well-defined means that we
can make firm experimental predictions.}.

\item Some hybrids are believed to be very {\it stable} (i.e.\ having small
widths). For hybrids below the $\dss D$ threshold the flux tube
model predicts very small widths (decays into $D D$, $\ds\ D$ and
$\ds\ \ds$ are almost forbidden). Hence we are specifically interested in
hybrid decay widths to $D^{\ast
\ast} D$ above threshold.

\end{itemlist}

The reason for our interest in {\it charmonium} hybrids derives from
the expectation\cite{isgur,paton} that their masses are better defined
than is the case for their light quark counterparts and, due to the smaller
amount of phase space available in the corresponding decay channels, their
widths are smaller. Bottomonium hybrids
are expected to be more difficult to produce than charmonium hybrids.
Experimentally, the
situation is

\begin{romanlist}

\item {\it Light quark hybrids :} There has recently been a claim of a broad
$J^{PC} = 1^{-+}$ exotic resonance with isospin one by the AGS at
Brookhaven\cite{brookhaven}, with mass in the
range 1.6 -- 2.2 GeV. They studied pion-nucleus inelastic scattering,
and found evidence for the above resonance decaying into $f_{1}(1285)
\pi^{-} \rightarrow K^{+} \pi^{0} \pi^{-}$, a process which is
predicted to be important (with a width of 50 MeV) in the flux tube
model\cite{paton}. The current data suffers from low statistics,
making its discovery an ambiguous claim and its existence in need of
independent corroboration.

\item {\it Charmonium hybrids :} Beijing $e^{+} e^{-}$ annihilation experiments
may soon have high enough luminosity to produce $J^{PC} = 1^{--}$
hybrids above the $D D$ threshold.

\item {\it Bottomonium hybrids :} Detection in $e^{+} e^{-}$
annihilation of a $J^{PC} = 1^{--}$ hybrid above the $B B$ threshold
at the SLAC B-factory is a possibility.

\end{romanlist}

\vspace{-0.4cm}
\section{The flux-tube model decay amplitude for $hybrid \rightarrow
\dss D$}
\vspace{-0.2cm}

In the Isgur-Paton non-relativistic flux-tube model of QCD\cite{isgur}
the {\it gluonic field} of a hybrid is represented by {\it beads},
which are connected
to each other and the quarks at the ends via a non-relativistic string.
The gluonic field is excited, giving rise to an excited adiabatic potential
between the quarks, which enables the study of hybrids without
further experimental input.

The decay amplitude\cite{paton,kokoski} of $A \rightarrow B C$ by pair
creation is similar to the $^{3}P_{0}$-model amplitude (where $A$ is
the initial meson decaying into mesons $B$ and $C$).
There is an additional {\it overlap} of the string of A to break at the
pair creation position into the strings of B and C, though. The model also
predicts an {\it overlap} for $hybrid \rightarrow B C$. This prohibits pair
creation on the hybrid q\={q}-axis.

We perform an {\it analytical calculation}\cite{page} of the decay amplitude
$hybrid \rightarrow \dss D$ by assuming the outgoing \dss\ and $D$ wave
functions to be L=1 and L=0
S.H.O. wave functions with inverse radii $\beta_{\dss}$ and
$\beta_{D}$ respectively. The initial hybrid wave function is
proportional to $r^{\delta} D(\Omega) \exp (- {\beta_{hybrid}^{2}
r^{2}}/2)$, where $D(\Omega)$ is a Wigner rotation function and
$0<\delta <1$.

The calculation is performed for the case $\beta_{\dss} = \beta_{D}
\equiv \beta$. The main reason for this simplification is that \mbox{{\it (a)}
the} \dss, \ds\ and $D$ are expected to have similar $\beta$'s, and
that \mbox{{\it
(b)} the} systematics of earlier numerical calculations for light
quarks can be understood in this limit.

The only free parameter in the model is the {\it lattice size} (or
longitudinal distance between beads), which is related to the
overall normalization of decays\cite{kokoski}. All other parameters
have previously been estimated in the context of the
model\cite{isgur}.

For decays of {\it ordinary mesons into ordinary mesons}, it is generally the
case that the $^{3}P_{0}$ and the flux-tube models
coincide in the limit where the string tension vanishes. It is
possible to make a stronger statement : {\it The $^{3}P_{0}$ and
flux-tube models coincide when $\beta_{B} = \beta_{C}$ even if the
string tension is non-zero}\cite{page}. This result at least holds for two
final
state L=0 mesons {\it and} for final state L=0 and L=1 mesons. This also
explains
the systematics of earlier numerical calculations\cite{kokoski}.

For decays of {\it hybrid mesons into ordinary mesons}, the flux-tube model
predictions are much more distinctive. When $\beta_{B} = \beta_{C}$ the
hybrid decay width to two L=0 mesons is zero (i.e. $hybrid \rightarrow
D D, \ds D,$ and $\ds \ds$ is forbidden). Hence the first non-zero decay
width is to L=1 and L=0 mesons. This explains our interest in $hybrid
\rightarrow \dss D$.

When $\beta_{B} = \beta_{C}$ there is also an important {\it selection
rule} operating in the moving frame of the initial q\={q}-pair for $hybrid
\rightarrow \dss D$ : The one unit of angular momentum of the hybrid
around the q\={q}-axis is exactly absorbed by the component of the
outgoing \dss\ angular momentum along the q\={q}-axis.

The dominant decay modes of $hybrid \rightarrow \dss D$ are displayed in
Table \ref{hybriddecay}. The eight lowest lying hybrids in the model
are assumed to have masses of 4.30 GeV or 4.35 GeV (and in addition a
small hyperfine splitting\cite{merlin}, which affects phase space
appreciably). They lie around the $\dss D$ threshold
of $\approx 4.3$ GeV. The magnitudes of the decays are
normalized to ordinary meson decays\cite{kokoski}. All resonances are
approximated to be narrow. We assume the $D$, $D^{\ast \ast}_{2^{++}}$,
$D^{\ast
\ast}_{1^{+L}}$ (low mass), $D^{\ast \ast}_{0^{++}}$ and $D^{\ast
\ast}_{1^{+H}}$ (high mass) have masses 1.87, 2.46, 2.42, 2.40 and
2.45 GeV, respectively. Also $\beta_{hybrid}$
= 0.35 GeV, $\beta$ = 0.37 GeV, $\delta = 0.62$ and the $D^{\ast
\ast}_{1^{+L}}$ / $D^{\ast
\ast}_{1^{+H}}$ mixing is $41^{o}$.

\baselineskip=12pt
\begin{table}[h]
\begin{center}
\tcaption{Dominant widths in MeV for {\it hybrid (c\={c}g)} $\rightarrow\ \dss
D$
for various $J^{PC}$ in partial wave {\it L}, both for 4.30 GeV hybrids
($\Gamma_{1}$) and 4.35 GeV hybrids ($\Gamma_{2}$).}
\label{hybriddecay}
\begin{tabular}{|l|l|c|r|r||l|l|c|r|r||l|l|c|r|r|}
\hline %------------------------
\it{c\={c}g}   & \dss     & \it{L} & $\Gamma_{1}$ & $\Gamma_{2}$ &
\it{c\={c}g}   & \dss     & \it{L} & $\Gamma_{1}$ & $\Gamma_{2}$ &
\it{c\={c}g}   & \dss     & \it{L} & $\Gamma_{1}$ & $\Gamma_{2}$ \\
\hline \hline %------------------------
$2^{-+}$ & $2^{++}$ & S & 0    & 120  & $2^{+-}$ & $2^{++}$ & P & 0&
40      & $1^{++}$ & $2^{++}$ & P & 0    & 40  \\
%------------------------
         & $1^{+L}$ & D & .02 &  10  &          & $1^{+L}$ & P & .3
& 10      &          & $1^{+L}$ & P & 6    & 70  \\
%------------------------
         & $0^{++}$ & D & .9  &   8  &          & $1^{+H}$ & P & 0
& 40      &          & $1^{+H}$ & P & 0    & 50  \\
\hline %------------------------
$1^{-+}$ & $1^{+L}$ & S &20    &  30  & $1^{+-}$ & $2^{++}$ & P & 0
& 20      &$0^{+-}$ & $1^{+L}$ & P &30    &150  \\
%------------------------
         &          & D & 1    &  40  &          & $1^{+L}$ & P &20
&120      &          & $1^{+H}$ & P & 0    & 90  \\
\cline{11-15} %------------------------
         & $1^{+H}$ & S & 0    & 300  &          & $0^{++}$ & P &50
&140      &$1^{--}$  & $1^{+L}$ & S &60    &150  \\
\cline{1-5} %------------------------
$0^{-+}$ & $0^{++}$ & S & -    & 400  &          & $1^{+H}$ & P & 0
& 30      &          & $1^{+H}$ & S & 0    &170  \\
\hline %------------------------
\end{tabular}
\end{center}
\end{table}

\vspace{-0.4cm}
\section{Conclusions and Acknowledgements}
\vspace{-0.2cm}
The  flux-tube model predicts a surprising {\it stability of gluonic
excitations} in heavy mesons. Above the $\dss D$ theshold hybrids
decay preferentially into $\dss D$ (with a width that is driven by the
available phase space), instead of $D D, \ds D, \ds \ds$, perhaps
explaining why hybrids have not yet been found experimentally.

I would like to thank Ted Barnes and Jack Paton for consultation.

\end{document}